\documentclass[preprint,floatfix] {revtex4} 
 
\usepackage{graphicx}
\usepackage{subfigure}
\begin{document}

\title{Ro-vibrational spectroscopy of molecules represented by a Tietz-Hua oscillator potential}
\author{Amlan K.~Roy}
\altaffiliation{Email: akroy@iiserkol.ac.in, akroy6k@gmail.com}
\affiliation{Division of Chemical Sciences,   
Indian Institute of Science Education and Research (IISER)-Kolkata, 
Mohanpur Campus, P. O. BCKV Campus Main Office, Nadia, 741252, WB, India}

\begin{abstract}
Accurate low and high-lying bound states of Tietz-Hua oscillator potential are presented. The radial Schr\"odinger equation
is solved efficiently by means of the generalized pseudospectral method that enables optimal spatial discretization. Both $\ell=0$ 
and rotational states are considered. Ro-vibrational levels of six diatomic molecules \emph{viz.}, H$_2$, HF, N$_2$, NO, O$_2$, 
O$_2^+$ are obtained with good accuracy. Most of the states are reported here for the first time. A detailed analysis of variation 
of eigenvalues with $n, \ell$ quantum numbers is made. 
Results are compared with literature data, wherever possible. These are also briefly contrasted with the Morse potential results.

\end{abstract}
\maketitle

\section{Introduction}
Construction of the universal potential energy function for molecules has been a challenging and active field of
research in chemical physics. The reason for this is that the potential energy function succinctly carries the 
necessary informations relevant for a molecule. Thus, an enormous number of such functions have been proposed over the
years, after the publication of three-parameter, exponential Morse potential \cite{morse29} about 85 years ago.  
The literature is huge; the following reference gives some of the older as well as relatively newer empirical 
functions \cite{morse29,rydberg31,rosen32,poschl33,manning33,lippincott53, frost54,tietz63,murrell74,schioberg86,hua90,
zavitsas91,rafi95,gardner99,hajigeorgiou2000,ikhdair07,ikhdair08,hajigeorgiou2010}. Usually, larger the number of parameters in 
the analytical potential energy function, better the fit with experimental data. While a few of these such as Morse, Mie-type
and pseudoharmonic potentials offer \emph{exact} analytic solutions \cite{ikhdair07,ikhdair08}, most of these unfortunately 
can not be analytically solved for arbitrary vibrational and rotational quantum numbers. This necessitates the use of 
approximation schemes for their solutions. 

Recently some attention has been paid on an analytic Tietz-Hua (TH) model potential \cite{tietz63,hua90} for
ro-vibrational levels in diatomic molecules, expressed in the following form, 
\begin{equation}
v(r) = D \left[ \frac{1-e^{-b_h(r-r_e)}}{1-c_he^{-b_h(r-r_e)}} \right]^2; \ \ \ b_h=\beta (1-c_h),
\end{equation} 
where $r_e$ relates to the molecular bond length, $\beta$ the Morse constant, $D$ signifies the potential well depth, $r$ denotes
the internuclear distance, while $c_h$ implies an optimization parameter obtained from \emph{ab initio} or Rydberg-Klein-Rees 
intramolecular potentials respectively. Note that in the limit of potential constant $c_h$ approaching 
zero, TH potential reduces to the familiar Morse potential \cite{morse29}. This potential supposedly describes the 
molecular dynamics (especially at high rotational and vibrational quantum numbers) more realistically than the traditional 
Morse potential \cite{pack72,levin90,natanson91}. Also it has been noted that this usually fits the experimental 
Rydberg-Klein-Rees 
curve more closely than the Morse function, especially near the dissociation limit \cite{pack72,levin90,hua90,natanson91}. 
In another study, using Hamilton-Jacobi theory in conjunction with Bohr-Sommerfeld quantization rule, analytical 
expressions for rotational-vibrational levels of diatomic molecules within TH model have been derived \cite{kunc97}. Radial 
probability distributions of some diatomic molecules in excited rotational-vibrational states have also been reported using this
route \cite{vazquez98}. Very recently, \emph{exact} analytical solution of the radial Schr\"odinger equation with TH potential has
been provided for $s$ waves within a parametric Nikiforov-Uvarov method \cite{hamzavi2012}. In another development, approximate
analytical solutions of the Dirac equation with TH potential were obtained for arbitrary spin-orbit quantum number using
the Pekeris scheme \cite{ikhdair2012}.  

The purpose of this work is to offer approximate solution of radial Schr\"odinger equation with TH potential for molecules.
As already mentioned, the $\ell=0$ states of this potential can be obtained in closed analytic form; while eigenvalues and 
eigenfunctions of $\ell \ne 0$ states of TH oscillator has not yet been reported in the literature, to the best of our 
knowledge. Here we take the help of generalized pseudospectral method (GPS) for an optimal effective discretization of the 
relevant Schr\"odinger equation. This method has produced very promising results for a number of situations having 
physical, chemical interest, including structure, dynamics in atomic and molecular physics. Accurate eigenvalues, eigenfunctions
were obtained for low as well as higher states for a class of potentials such as spiked harmonic oscillators, logarithmic, 
rational, power-law, Hulth\'en, Yukawa, exponentially screened coulomb potentials, etc. \cite{roy04a,roy05,roy05a, 
roy08a,roy13,roy13a}. Thus we make a detailed study on the bound-state spectrum of TH oscillator with particular 
reference to diatomic molecules. Ro-vibrational energies and radial densities are studied for both $s$-wave and \emph{rotational} 
sates having \emph{arbitrary} low and high vibrational quantum number. This will also enable us to judge the viability and feasibility 
of current approach in the context of diatomic molecular potentials. To this end, arbitrary $\{n,\ell\}$ states are reported for six 
molecules, \emph{viz.}, H$_2$, HF, N$_2$, NO, O$_2$, O$_2^+$. Comparison with literature data are made wherever possible. The article is
organized as follows. A brief overview of the adopted method is given in Section II. Then a discussion of the results is presented in
Section III, while we conclude with a few remarks in Section IV. 

\section{The GPS method}
This method has been discussed in detail earlier (see the references \cite{roy04a,roy05,roy05a,roy08a,
roy13,roy13a} and therein). Thus it suffices to present here only a brief summary of the essential steps involved.  

Without loss of generality, the desired time-independent radial Schr\"odinger equation, to be solved, can be written 
as (atomic units employed unless otherwise mentioned),
\begin{equation}
\left[-\frac{1}{2} \ \frac{d^2}{dr^2} + \frac{\ell (\ell+1)} {2r^2} + v(r) \right] \psi_{n,\ell}(r)=E_{n,\ell}\ \psi_{n,\ell}(r)
\end{equation}
where $v(r)$ is the TH potential, as given in Eq.~(1), while $n$, $\ell$ signify the radial and angular momentum quantum numbers 
respectively. The GPS formalism facilitates the use of a denser mesh at small distance and relatively coarser mesh at large
distance preserving similar accuracy at both the regions.

A key step in this approach is to approximate a function $f(x)$ defined in the interval $x \in [-1,1]$ by an N-th order polynomial 
$f_N(x)$ {\it exactly} at the discrete collocation points $x_j$, 
\begin{equation}
f(x) \cong f_N(x) = \sum_{j=0}^{N} f(x_j)\ g_j(x), \ \ \ \ \ \ \ \ \  f_N(x_j) = f(x_j).
\end{equation}
Within the Legendre pseudospectral method that we are using currently, $x_0=-1$, $x_N=1$, and $x_j$'s $(j=1,\ldots,N-1)$ are determined 
from the roots of first derivative of the Legendre polynomial $P_N(x)$ with respect to $x$, i.e., $P'_N(x_j) = 0$. The $g_j(x)$s in 
Eq.~(3) are termed the cardinal functions expressed as,
\begin{equation}
g_j(x) = -\frac{1}{N(N+1)P_N(x_j)}\ \  \frac{(1-x^2)\ P'_N(x)}{x-x_j},
\end{equation}
satisfying the relation $g_j(x_{j'}) = \delta_{j'j}$. At this stage, a transformation $r=r(x)$ is used to map the semi-infinite domain  
$r \in [0, \infty]$ onto the finite domain $x \in [-1,1]$, along with an algebraic nonlinear mapping, 
$r=r(x)=L\ \ \frac{1+x}{1-x+\alpha}$, with L, $\alpha=2L/r_{max}$ being two mapping parameters. Finally introducing a symmetrization 
procedure gives a transformed Hamiltonian of the following form, 
\begin{equation}
\hat{H}(x)= -\frac{1}{2} \ \frac{1}{r'(x)}\ \frac{d^2}{dx^2} \ \frac{1}{r'(x)} + v(r(x))+v_m(x).
\end{equation}
The advantage lies in the fact that this leads to a \emph{symmetric} matrix eigenvalue problem which can be solved readily and 
efficiently to give accurate eigenvalues, eigenfunctions by using standard routines. Note that $v_m(x)=0$ for the above 
transformation and one finally obtains a set of discretized coupled equations.

Considerable checks have been made on the convergence of eigenvalues with respect to the mapping parameters for a decent number of 
molecular states. After a series of such test calculations, a choice has been made at the point where the results changed 
negligibly with such variations. In this way, a consistent and uniform set of parameters ($r_{max}=500,$ $\alpha=1$ and $N=300$) 
has been used. 

\section{Results and Discussion}
\begingroup
\squeezetable
\begin{table}
\caption {\label{tab:table1}Spectroscopic parameters of the molecules, used in present calculation, taken from 
\cite{kunc97}.}
\begin{ruledtabular}
\begin{tabular}{ccccccc}
Molecule  &  $c_h$     &  $\mu/10^{-23}$ (g) &  $b_h (\AA^{-1})$ & $r_c (\AA^{-1})$ &  $\beta (\AA^{-1})$  &  $D_e$(cm$^{-1}$) \\
\hline
H$_2$     &  0.170066    & 0.084               & 1.61890           & 0.741            &  1.9506              & 38318       \\ 
HF        &  0.127772    & 0.160               & 1.94207           & 0.917            &  2.2266              & 49382       \\ 
N$_2$     & $-$0.032325  & 1.171               & 2.78585           & 1.097            &  2.6986              & 79885       \\ 
NO        & $-$0.029477  & 1.249               & 2.71559           & 1.151            &  2.7534              & 53341       \\
O$_2$     &  0.027262    & 1.377               & 2.59103           & 1.207            &  2.6636              & 42041       \\
O$_2^+$   & $-$0.019445  & 1.377               & 2.86987           & 1.116            &  2.8151              & 54688       \\ 
\end{tabular}
\end{ruledtabular}
\end{table}
\endgroup

At first, we present the calculated ro-vibrational levels within the TH model potential. For this six representative molecules, \emph{viz.},
H$_2$, HF, N$_2$, NO, O$_2$, O$_2^+$ are selected; the respective spectroscopic parameters, adopted from \cite{kunc97}, are given in Table I. 
The conversion parameters used in this work are taken from NIST database \cite{nist}. These are as follows: Bohr radius = 0.52917721092 \AA, 
Hartree energy = 27.21138505 eV, and electron rest mass = 5.48577990946 $\times 10^{-4}$ u. However, before we proceed, note that this 
potential reduces to Rosen-Morse, Morse and Manning-Rosen potential for negative, zero and positive values of $c_h$ respectively 
\cite{kunc97}. Thus two sets of calculations are performed for H$_2$ and HF in the limit of $c_h=0$. The $\{0,0\}, \{5,0\}, \{7,0\}$ state
energies at such limit are estimated to be $-$4.481469 ($-$4.481466), $-$2.220206 ($-$2.220195), $-$1.535804 ($-$1.535780) eV, where the 
numbers in the parentheses refer to similar energies reported in \cite{hamzavi2012}, obtained by means of a Nikiforov-Uvarov method. The
first and second integer in square bracket identify the vibrational (radial) and rotational (angular) quantum numbers respectively. 
The same three states for HF molecule read as follows $-$5.868677 ($-$5.868710), $-$3.625307 ($-$3.625604) and $-$2.878718 ($-$2.878878). 
In both cases, our GPS results are found to be in excellent agreement with those from \cite{hamzavi2012} and one notices that when the 
potential constant $c_h$ tends to zero, TH energy levels approach that of the familiar Morse oscillator levels. This has been verified for
other states as well. Now we present the main results for $s$-wave and rotational states for these six molecules in Table II. Thus, 
nine low-lying bound-state energies corresponding to $\{0,\ell\}, \{3,\ell\}, \{5,\ell\}$, having $\ell=0,1,2$ are reported. In comparison 
to other molecular potentials, there is a visible lack of literature results for TH oscillator potential. No direct results are available for
any of the non-zero angular momentum states. Only the $\ell=0$ states having vibrational quantum number $n=0,5,7$ have been reported very 
recently in a parametrically generalized Nikiforov-Uvarov formalism \cite{hamzavi2012}. These are available for all the five molecular 
species except NO. In all ten occasions, GPS energies are found to be in very good agreement with the literature values. The slight 
discrepancy may be due to the slight differences in conversion factors used in \cite{hamzavi2012}. Next, in Table III, nine high-lying 
ro-vibrational energies are reported for all the 6 molecular species. Angular quantum number as high as $\ell=30$ is considered. To the best
of our knowledge, none of these states have been reported before and it is hoped that these could be useful for future referencing. 

\begingroup
\squeezetable
\begin{table}
\caption {\label{tab:table2}Calculated eigenvalues of TH potential for some low-lying 
states of six diatomic molecules along with literature data. PR signifies Present Result.} 
\begin{ruledtabular}
\begin{tabular}{cc|ll|ll|ll}
$n$  &  $\ell$  & \multicolumn{2}{c}{E$_{n,\ell}-$D$_e$ (eV)} & \multicolumn{2}{c}{E$_{n,\ell}-$D$_e$ (eV)} & \multicolumn{2}{c}{E$_{n,\ell}-$D$_e$ (eV)} \\
\hline
  &   &    PR   &   Literature    &  PR    & Literature     &   PR   & Literature  \\
\hline
   &      &   \underline{H$_2$}   &   &    \underline{HF}    &  &   \underline{N$_2$}      &    \\
  0 &  0  & $-$4.4815797825  & $-$4.481571826 & $-$5.8687195228  & $-$5.868757846  & $-$9.7588058322  & $-$9.7588029855  \\
  3 &     & $-$3.0595425362  &                & $-$4.4737571516  &                 & $-$8.9066675119  &                  \\
  5 &     & $-$2.2815913849  & $-$2.281533873 & $-$3.6601740988  & $-$3.660498629  & $-$8.3595761578  & $-$8.359551147   \\
  0 &  1  & $-$4.4669801579  &                & $-$5.8636625262  &                 & $-$9.7583155848  &                  \\
  3 &     & $-$3.0474413866  &                & $-$4.4692935886  &                 & $-$8.9061913507  &                  \\
  5 &     & $-$2.2710928924  &                & $-$3.6560952745  &                 & $-$8.3591095263  &                  \\
  0 &  2  & $-$4.4379154622  &                & $-$5.8535547327  &                 & $-$9.7573351069  &                  \\
  3 &     & $-$3.0233638406  &                & $-$4.4603723647  &                 & $-$8.9052390449  &                  \\
  5 &     & $-$2.2502130058  &                & $-$3.6479433575  &                 & $-$8.3581762808  &                  \\
    &     &   \underline{NO}      &   &  \underline{O$_2$}   &  &    \underline{O$_2^+$}  &    \\
  0 &  0  & $-$6.4959334209 &     & $-$5.1163223113  &  $-$5.116333496  & $-$6.6645714733  &  $-$6.6645687718  \\
  3 &     & $-$5.8133374461 &     & $-$4.5590745476  &                  & $-$5.9898565742  &                   \\
  5 &     & $-$5.3795826206 &     & $-$4.2058686976  & $-$4.205982010   & $-$5.5597008912  &  $-$5.559676435   \\
  0 &  1  & $-$6.4955164040 &     & $-$5.1159784440  &                  & $-$6.6641689327  &                   \\
  3 &     & $-$5.8129354009 &     & $-$4.5587436240  &                  & $-$5.9894673742  &                   \\
  5 &     & $-$5.3791906688 &     & $-$4.2055464879  &                  & $-$5.5593207256  &                   \\
  0 &  2  & $-$6.4946823862 &     & $-$5.1152907228  &                  & $-$6.6633638662  &                   \\
  3 &     & $-$5.8121313267 &     & $-$4.5580817907  &                  & $-$5.9886889893  &                   \\
  5 &     & $-$5.3784067819 &     & $-$4.2049020823  &                  & $-$5.5585604098  &                   \\
\end{tabular}
\end{ruledtabular}
\end{table}
\endgroup

\begingroup
\squeezetable
\begin{table}
\caption {\label{tab:table3}Calculated eigenvalues of TH potential for some high-lying
states of six diatomic molecules. PR signifies Present Result.} 
\begin{ruledtabular}
\begin{tabular}{ccl|ccl|ccl}
$n$  &  $\ell$  & E$_{n,\ell}-$D$_e$ (eV) & $n$  &  $\ell$ &  E$_{n,\ell}-$D$_e$ (eV) &  $n$ & $\ell$ & E$_{n,\ell}-$D$_e$ (eV)  \\
\hline
   & \underline{H$_2$}        &  &    &  \underline{HF}    &  &    &  \underline{N$_2$}   &    \\
  0  &  15  &  $-$2.9921666314   &   0  &  15  & $-$5.2763995044 &  0   &  10  &  $-$9.7318505513    \\ 
  6  &      &  $-$0.9895676615   &  10  &      & $-$1.6462931164 &  3   &      &  $-$8.8804871129    \\
  9  &      &  $-$0.3970747424   &  15  &      & $-$0.6211626101 &  5   &      &  $-$8.3339200142    \\
  0  &  20  &  $-$2.1321131585   &   0  &  20  & $-$4.8508452082 &  0   &  15  &  $-$9.7000161732    \\
  6  &      &  $-$0.4836490402   &  10  &      & $-$1.3893569074 &  3   &      &  $-$8.8495689016    \\
  9  &      &  $-$0.0621445548   &  15  &      & $-$0.4466120874 &  5   &      &  $-$8.3036216647    \\
  0  &  25  &  $-$1.2522963255   &   0  &  30  & $-$3.7271155639 &  0   &  20  &  $-$9.6559768987    \\
  4  &      &  $-$0.3093944960   &  10  &      & $-$0.7370870990 &  3   &      &  $-$8.8067988793    \\
  5  &      &  $-$0.1506049691   &  15  &      & $-$0.0365890609 &  5   &      &  $-$8.2617104279    \\
   & \underline{NO}      &  &  & \underline{O$_2$}  &  &  &   \underline{O$_2^+$}   &    \\
  0  &  10  &  $-$6.4730053894  &   0  &   10  &   $-$5.0974162346  &   0  &   10  &  $-$6.6424390361  \\
  3  &      &  $-$5.7912330333  &   3  &       &   $-$4.5408805335  &   3  &       &  $-$5.9684580465  \\
  5  &      &  $-$5.3580334807  &   5  &       &   $-$4.1881540596  &   5  &       &  $-$5.5387993854  \\ 
  0  &  15  &  $-$6.4459293673  &   0  &   15  &   $-$5.0750900921  &   0  &   15  &  $-$6.6163016863  \\
  3  &      &  $-$5.7651308474  &   3  &       &   $-$4.5193963303  &   3  &       &  $-$5.9431885057  \\ 
  5  &      &  $-$5.3325878765  &   5  &       &   $-$4.1672367007  &   5  &       &  $-$5.5141175736  \\
  0  &  20  &  $-$6.4084765984  &   0  &   20  &   $-$5.0442081090  &   0  &   20  &  $-$6.5801457931  \\
  3  &      &  $-$5.7290273103  &   3  &       &   $-$4.4896808485  &   3  &       &  $-$5.9082350230  \\
  5  &      &  $-$5.2973940820  &   5  &       &   $-$4.1383066355  &   5  &       &  $-$5.4799784798  \\
\end{tabular}
\end{ruledtabular}
\end{table}
\endgroup

Next we proceed for a detailed investigation on the energy variations for three selected molecules \emph{viz.,} H$_2$, HF and NO respectively.
The top three panels (a), (b), (c) in Fig.~1 depict the variations of E$_{n,\ell}-$D$_e$ (in eV) with respect to the angular quantum number 
$\ell$ for H$_2$, HF and NO. These are given for six values of vibrational quantum number, \emph{viz.,} $n=0,3,6,9,12,15$ for H$_2$; seven values 
of $n$ ($n=20$ in addition to all the six $n$ in H$_2$) for HF; and nine values of $n$ ($n=25,30$ in addition to all the seven $n$ in HF) for
NO. Note that the $\ell$ axis goes to 30 for H$_2$, while for the other two this is extended to 40. This happens because of the fact that
a limited number of bound states is supported by the potential. Such bound states occur in larger number for NO and HF than in H$_2$. It may
be worthwhile mentioning here that, the maximum vibrational quantum number $v_{\mathrm{max}}$ and maximum rotational quantum number
$\ell_{\mathrm{max}}$ for H$_2$, HF, N$_2$, NO, O$_2$, O$_2^+$ are (22,39), (28,66), (66,260), (56,230), (52,220) and (56,235) respectively
\cite{kunc97}. The above $v_{\mathrm{max}}$'s are to be contrasted with the corresponding values of 18, 23, 82, 67, 65 and 58 in Morse 
potential for the six molecules under investigation \cite{kunc97}. Much larger differences in $v_{\mathrm{max}}$ in TH potential (96) and Morse 
potential (174) have been observed in I$_2$, where the actual value is 107 \cite{appadoo96}. The plots for other three molecules are omitted, 
as their qualitative characteristic features remain similar to one of the three plots in (a), (b), (c). As one moves along the H$_2$-HF-NO 
series, the plot for a given $n$ series tends to vary rather slowly (rate of increase slows down), with H$_2$ and NO showing maximum and 
minimum increase respectively, such that for NO the plots are quite flat. Also for a given molecule, as one goes to higher $n$ values, the 
separation between two successive $n$ plots tends to decrease. Similar qualitative feature has been recorded earlier in the energy versus 
$\ell$ plot for H$_2$ within a semiclassical approach \cite{kunc97}. Now we turn to the E$_{n,\ell}-$D$_e$ (in eV) versus $n$ for fixed 
$\ell$ quantum number for the same three molecules in the bottom panels (d), (e) and (f) respectively. The $n$ axis in H$_2$, HF, NO extends 
to 25, 30 and 40. For H$_2$, these are studied at seven $\ell$, \emph{viz.,} 0,5,10,15,20,25,30, while for the other two molecules two more 
$\ell$ values of 35, 40 are considered besides the seven value of H$_2$. In going from H$_2$-HF-NO, the plots for different $\ell$ become 
progressively more closely spaced, with H$_2$ showing maximum sparsity and in NO the successive separations are too small to be identified 
properly in the scale. Another interesting feature is that, rate of increase in energy slowly increases as one moves the series, producing a 
nearly linear structure in NO. Also for H$_2$, HF, as $\ell$ takes higher values, the separation between to adjacent $\ell$ tends to grow 
large. In both the $\ell$ and $n$ plots in top and bottom panels, individual $n$ and $\ell$ series for a given molecule remain nearly parallel 
to each other.  

\begin{figure}
\centering
\begin{minipage}[c]{0.28\textwidth}\centering
\includegraphics[scale=0.30]{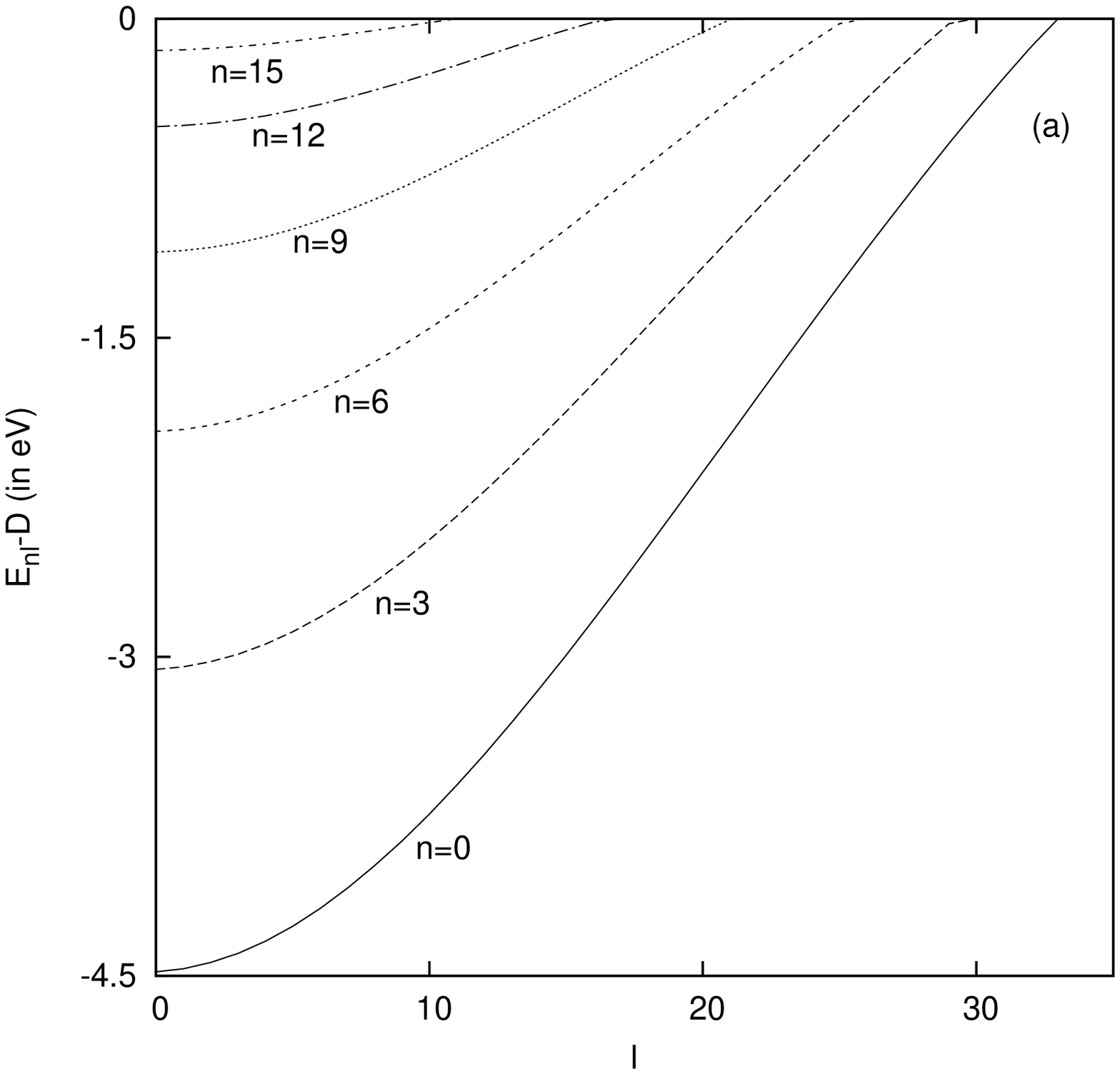}
\end{minipage}
\hspace{0.25in}
\begin{minipage}[c]{0.28\textwidth}\centering
\includegraphics[scale=0.30]{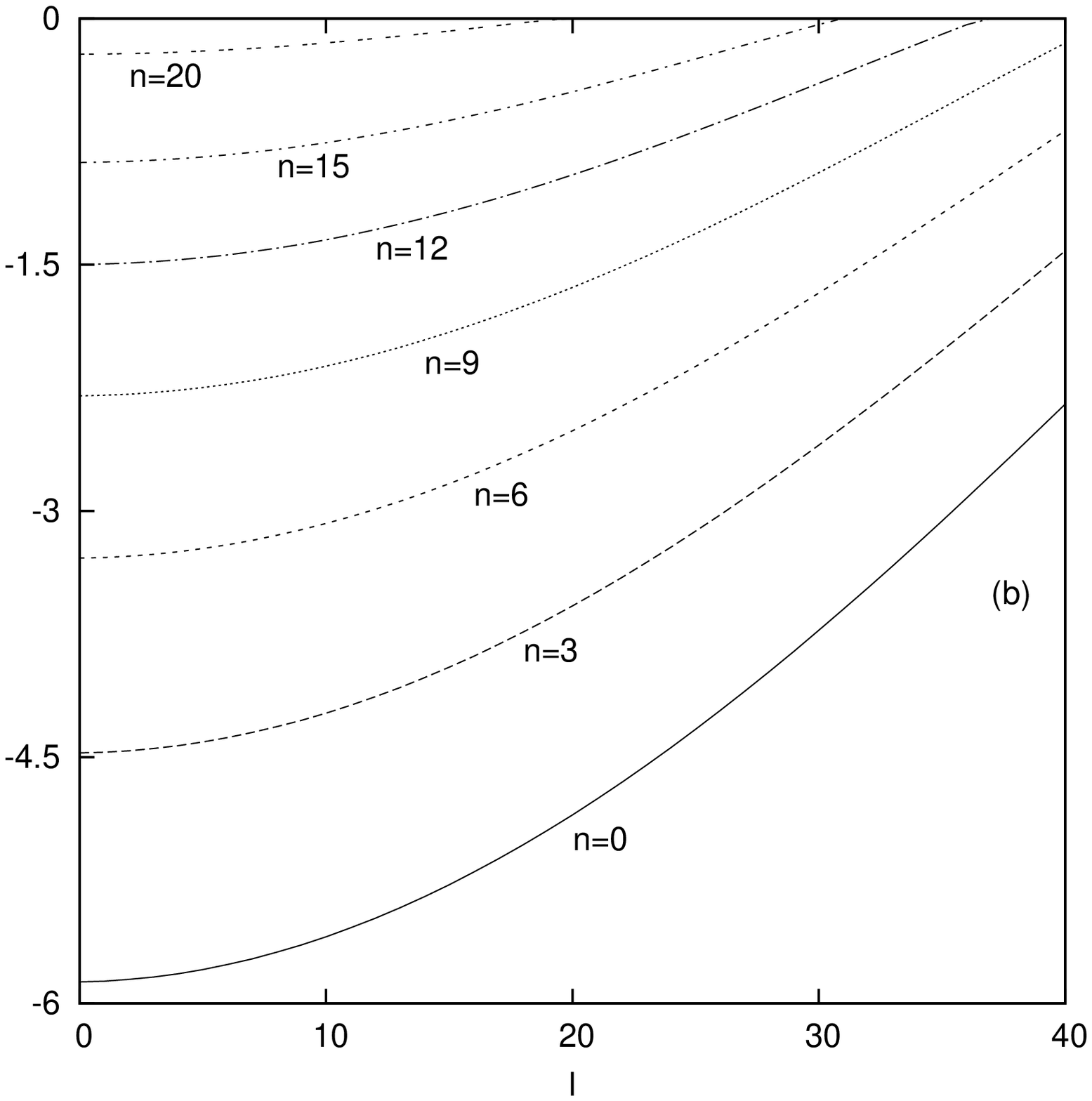}
\end{minipage}
\hspace{0.25in}
\begin{minipage}[c]{0.28\textwidth}\centering
\includegraphics[scale=0.30]{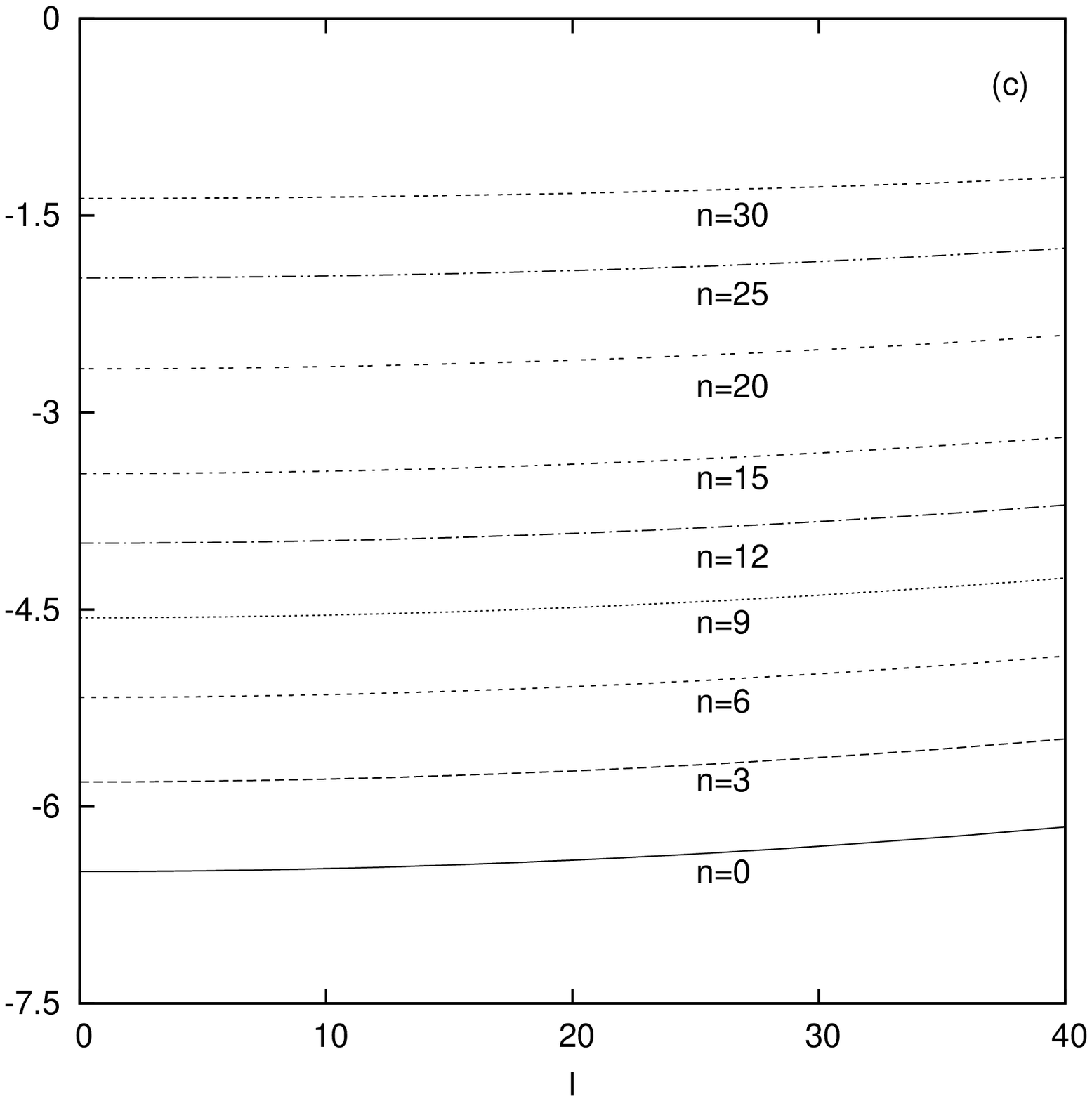}
\end{minipage}
\\[10pt]
\begin{minipage}[c]{0.28\textwidth}\centering
\includegraphics[scale=0.30]{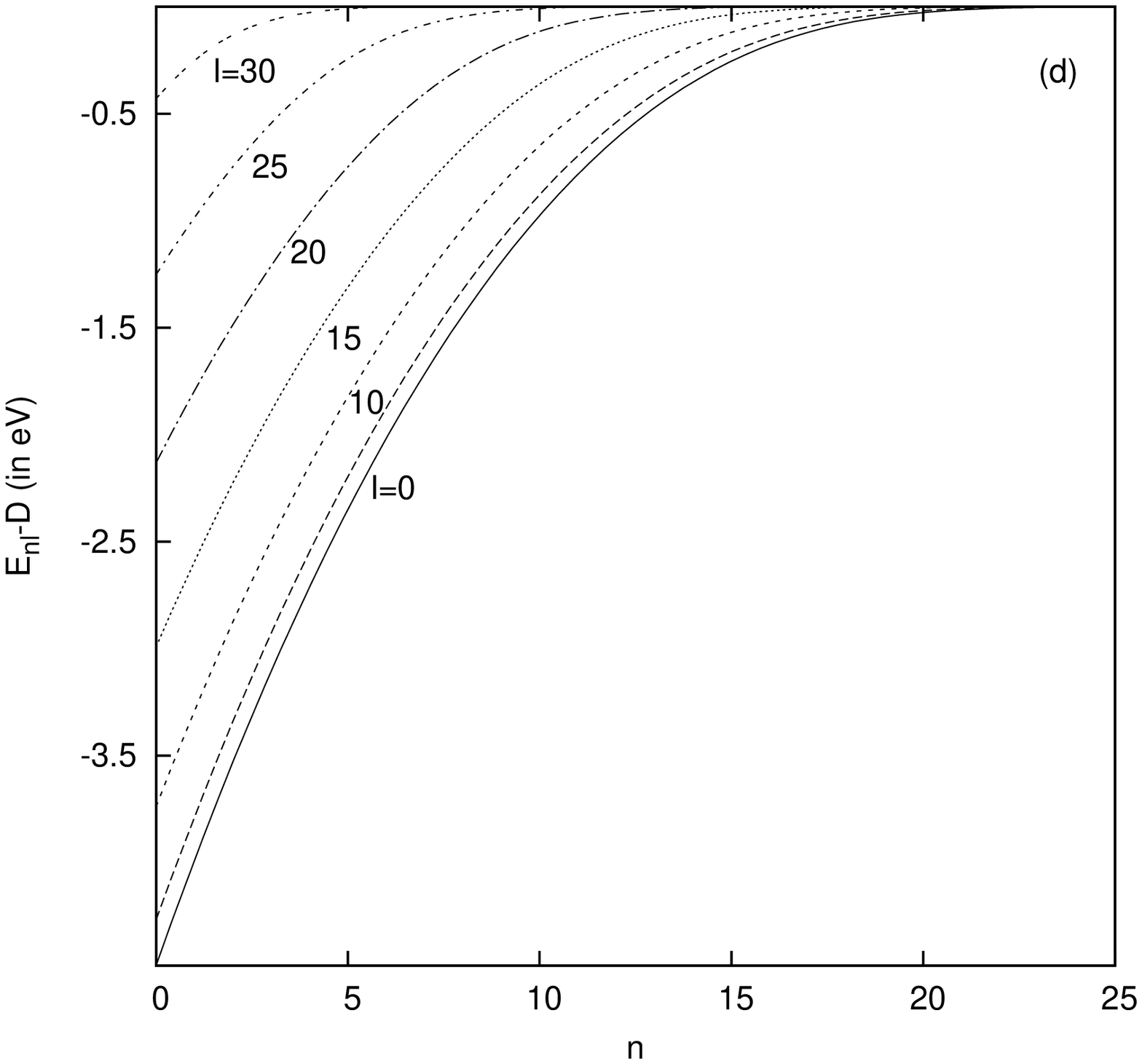}
\end{minipage}
\hspace{0.25in}
\begin{minipage}[c]{0.28\textwidth}\centering
\includegraphics[scale=0.30]{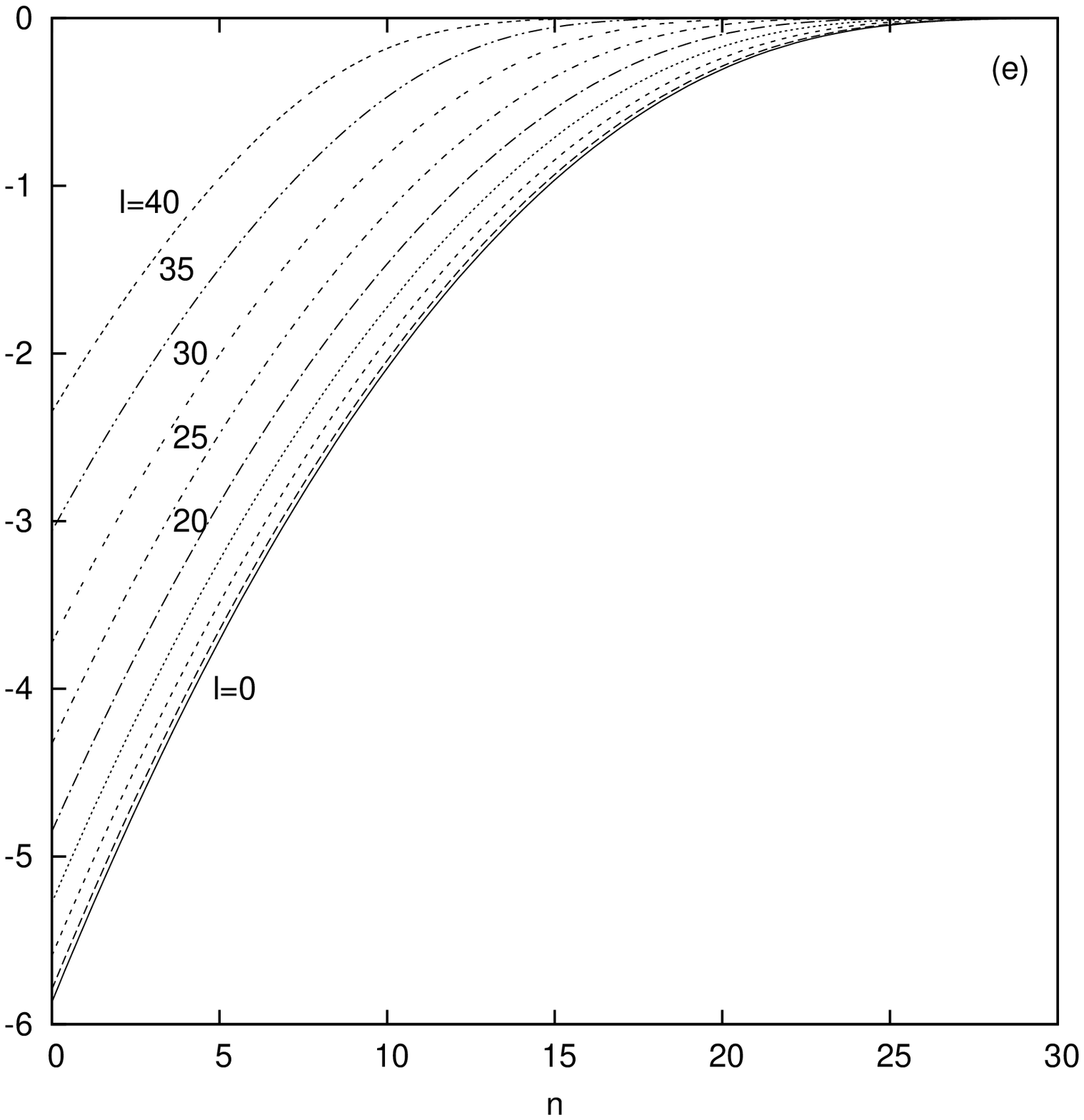}
\end{minipage}
\hspace{0.25in}
\begin{minipage}[c]{0.28\textwidth}\centering
\includegraphics[scale=0.30]{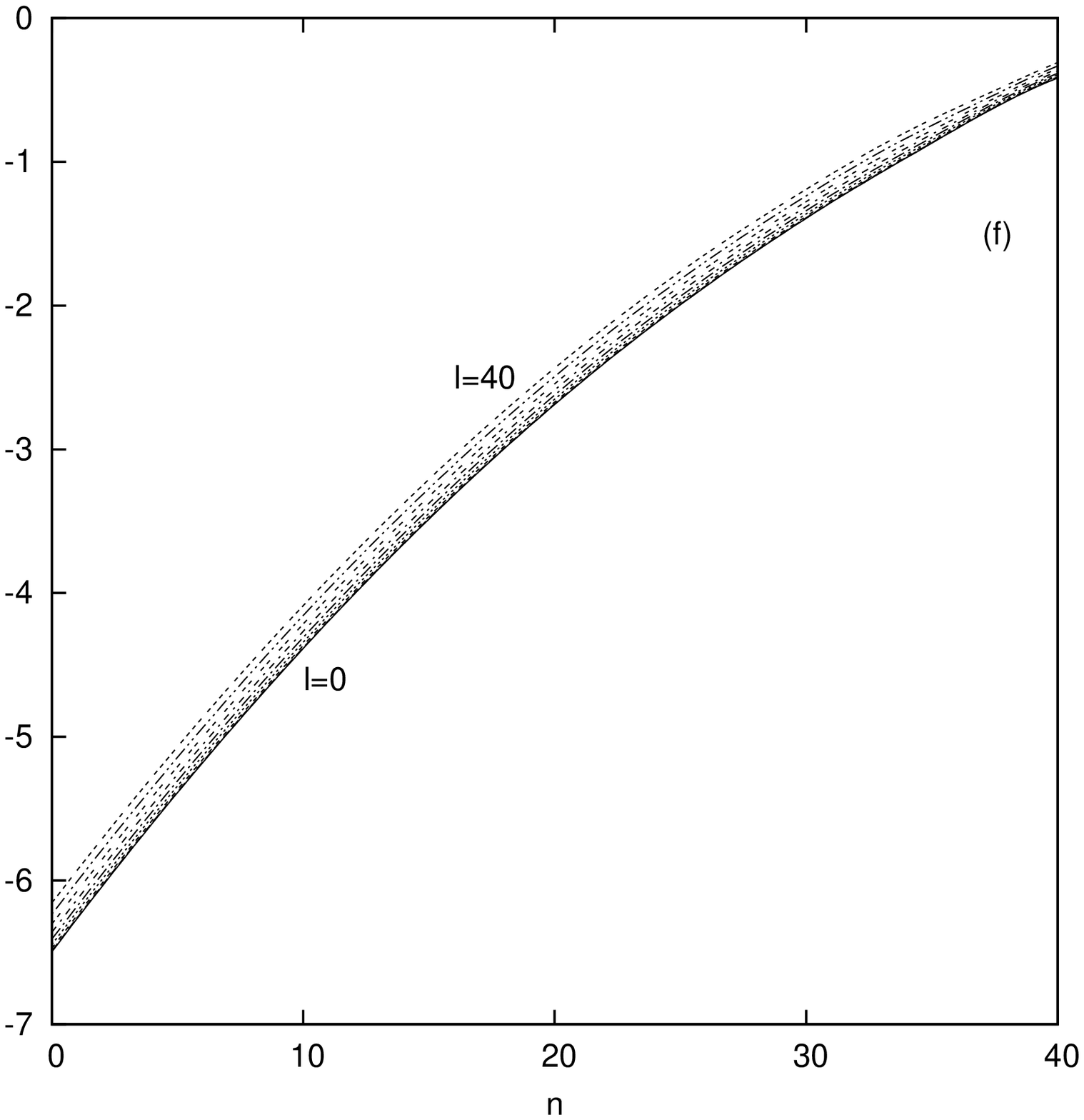}
\end{minipage}
\caption[optional]{Energy variations (in eV) in TH potential, with respect to rotational ($\ell$) (top panel) and vibrational ($n$) 
(bottom panel) quantum numbers respectively. The plots (a), (d) correspond to H$_2$, while (b), (e) and (c), (f) correspond to HF
and NO molecule respectively. State indices are indicated in the figure. See text for details.}
\end{figure}

\section{conclusion}
Tietz-Hua oscillator has been found to be a more realistic analytical potential than the familiar Morse potential in describing molecular 
dynamics at moderate as well as high rotational and vibrational quantum number. In the present work, we have presented both $s-$wave and 
rotational bound states having arbitrary rotational and vibrational quantum numbers with excellent accuracy. A total of 18 low and 
moderately high-lying ro-vibrational levels are given for six diatomic molecules, namely, H$_2$, HF, N$_2$, NO, O$_2$, O$_2^+$.
While the lower states match quite well the lone literature result, many new states are given here for the first 
time. Energy changes with respect to $n, \ell$ quantum numbers are discussed in detail for three molecules. In short, a simple accurate and 
efficient scheme is offered for this and other similar potentials in molecular physics.  


\end{document}